\documentclass[aps,prl,twocolumn,superscriptaddress,amsmath,amssymb,floatfix]{revtex4-2}

\usepackage{graphicx}
\usepackage{amssymb}
\usepackage{amsmath}
\usepackage{natbib}
\usepackage{color}
\usepackage{chngcntr}
\usepackage{enumitem}
\usepackage{hyperref}
\hypersetup{
colorlinks=false,
linktocpage=true,
citebordercolor=[rgb]{0.5,0.5,1.0},
linkbordercolor=[rgb]{1.0,0.5,0.5},
urlbordercolor=[rgb]{0.5,0.5,1.0}
}

\begin{document}

\title{Matching Rules for a Three-Dimensional Strongly Aperiodic Monotile}

\author{Felix Flicker}
\affiliation{School of Physics, University of Bristol, Bristol BS8 1TL, United Kingdom}

\date{\today}

\begin{abstract}
A recent pre-print~\cite{tsiokos} proposed a three-dimensional (3D) strongly aperiodic monotile: a shape that tiles Euclidean space only non-periodically and which admits no symmetry of infinite order. The proof takes the 3D Chair tile identified previously by Lee and Moody~\cite{LeeMoody2001}, and adds geometric decorations to the faces so as to force non-periodicity (without these decorations The Chair also admits periodic tilings). Here we establish general requirements on face decorations to achieve the same end, in order to facilitate the search for physical realisations. We find that the requirements are minimal. We provide matching rules using three colours of arrow that are equivalent to the original rules, in that they force the same local and global configurations. These rules force Chairs to compose into `Superchairs' with doubled linear dimensions. In this process the matching rules themselves compose uniquely. We find that the same global structure can be forced using simpler rules based on the colours of squares, regardless of orientation. Any physical system encoding these rules (geometrically or otherwise) will force the same strongly aperiodic monotilings. We provide simple examples.
\end{abstract}

\maketitle

2023 saw the discovery of the first two-dimensional (2D) aperiodic monotile: a shape that tiles the plane only non-periodically\footnote{We use the nomenclature that a tiling is `non-periodic' there is no vector through which a copy of the the tiling can be translated so as to perfectly overlap the original, and `aperiodic' if its tiles do not also admit any periodic tiling.}, meaning that there is no vector through which a copy of the the tiling can be translated so as to perfectly overlap the original~\cite{hat,spectre}. The story behind the discovery was both inspirational and enheartening, with the discoverer David Smith having pieced the initial tiling together by hand using cardboard shapes before formal proofs of its aperiodicity were provided by his collaborators. The discovery brought renewed urgency to the question of whether there might be a 3D version. In three dimensions a distinction is made between strongly and weakly aperiodic monotiles. A monotile is weakly aperiodic if it admits a tiling with a symmetry group of infinite order: for example, a screw axis along which a 2D tiling repeats periodically but with copies rotating through an incommensurate angle so as never to repeat. Strongly aperiodic monotiles admit only symmetries of finite order~\cite{hat,BaakeGrimm}. The Schmitt--Conway--Danzer biprism~\cite{Danzer1995} admits a screw axis, and the three-dimensional Socolar--Taylor tile~\cite{SocolarTaylor2011,SocolarTaylor2012} admits a periodic stacking of non-periodic layers, so both are weakly aperiodic. Aperiodic tilings requiring multiple tiles have been known since the 1960s~\cite{Berger1966,Robinson1971,GoodmanStrauss1999b}, and aperiodic sets of Wang cubes exist in three dimensions~\cite{CulikKari1995}. The question of whether there exists a strongly aperiodic 3D monotile had remained open.

\begin{figure}[t]
    \centering
    \includegraphics[width=0.9\columnwidth]{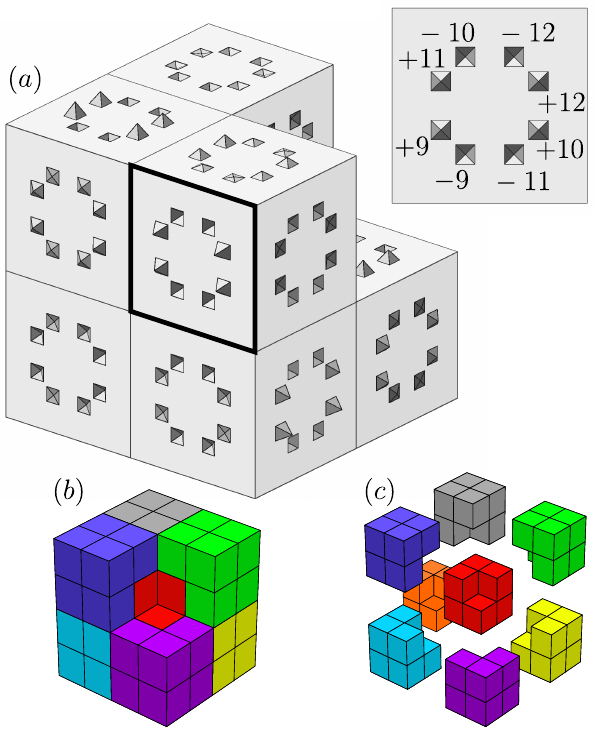}
\caption{(a) The 3D strongly aperiodic monotile identified in Ref.~\onlinecite{tsiokos}, referred to as Chair44. The inset shows the highlighted face, along with the heights or depths of the corresponding pyramid decorations that force aperiodicity in units of $1/10,000$ the cube length. Figure reproduced from Ref.~\onlinecite{tsiokos}, whose Fig.3 shows the patterns for all faces. (b) 8 Chairs combined into a Superchair. (c) The Superchair separated for clarity.}
    \label{fig:DC}
\end{figure}

A recent pre-print appears to have answered this question positively, using a method diametrically opposed to that employed in the 2D discovery: Ioannis Tsiokos provided the question to an Artificial Intelligence (AI, OpenAI's GPT-6 Astra), which proposed a 3D aperiodic monotile and which constructed a formal proof of aperiodicity written in the Lean proof-assistant programming language~\cite{tsiokos}. We verified that the Lean code reports success~\footnote{This does not mean the proof is correct. Craig Kaplan points out that there are known exploitable bugs in Lean. See \emph{e.g.} https://sechub.in/view/3288084.}. We have also independently reproduced the finite enumerations on which the proof rests (see below). The tile, named Chair44 in Ref.~\onlinecite{tsiokos} (Fig.~\ref{fig:DC})~\footnote{The name is presumably a reference to the 44 ways two Chair44 tiles can meet. However, only 30 of these can actually appear in the infinite tiling.}, starts from the 3D Chair tile, which is seven unit cubes forming a $2\times2\times2$ block with one corner removed. The Chair was known previously: Lee and Moody generalised the 2D Chair~\cite{BMS98} to every dimension, including 3D, showing that it admits infinite hierarchical non-periodic tilings (\emph{Chair Tilings})~\cite{LeeMoody2001}. (There are an uncountable infinity of locally isomorphic but globally distinct tilings.) However, The Chair can also tile periodically~\cite{Stein1990,Schmerl1994}. Chair44 is a decoration of the faces of The Chair with geometric features (square pyramids of different heights, and corresponding recesses) that aim to eliminate all but the non-periodic tilings, thus delivering aperiodic monotilings. These decorations are a form of matching rule, dictating how tiles must meet. For Chair44 to be an aperiodic monotile it is necessary to encode the matching rules geometrically, although we will find it convenient to temporarily abstract them.

The proof in Ref.~\onlinecite{tsiokos} then appears to proceed as follows~\footnote{C.~Goodman-Strauss has provided another human-readable proof~\cite{GoodmanStrauss}; our pre-prints appeared in the same arXiv listing, and we agree that our approaches are the same.}. First, enumerate all possible ways Chair44 can meet translated, possibly rotated, possibly mirrored, copies of itself: there are 44. Next, enumerate all possible ways to completely enclose Chair44 with copies of itself. There are 33 such clusters. Next, enumerate all ways to add a second shell around the central Chair44. This last check reveals that the central Chair44 can only ever appear as a member of one particular 8-Chair cluster, which we call The Superchair44 (Fig.~\ref{fig:DC}b,c). In one of the eight positions it sits at the heart of the cluster (red), touching all seven others. Without decorations, The Superchair is identical to The Chair with all lengths doubled~\cite{LeeMoody2001}. The second-shell enumeration also reveals that the 7 other Chairs in the Superchair can each be uniquely identified (Fig.~\ref{fig:DC}c): any Chair44 in an infinite Chair Tiling belongs to exactly one Superchair44.

The final step is to notice that the enumeration reveals that The Superchair's decorations force exactly the same matching rules for Superchairs as were present for Chairs. Hence the only valid infinite tilings of Chair44 must be hierarchical: they must group into Superchairs with the same structure on a doubled lengthscale, which must group into Super-Superchairs with the same structure on a quadrupled lengthscale, and so on. This provides a well known proof of aperiodicity, originating with the earliest works on aperiodic tilings by Berger~\cite{Berger1966,Robinson1971,GoodmanStrauss1998}: if the tiling maps onto itself under translation through some vector $\mathbf{v}$, then it must also map onto itself under translations through all vectors $2^{n}\mathbf{v}$ with $n$ any positive integer. The only vector for which $2^{n}\mathbf{v}\equiv\mathbf{v}$ for all positive integer $n$ is the zero vector. Hence, all infinite tilings are aperiodic. The same reasoning also bounds all symmetries: the tiling has at most the 24 symmetries of the cube. None is of infinite order, and so the tiling is strongly aperiodic.

Here we ask which aspects of the geometrical matching rules identified in Ref.~\onlinecite{tsiokos} are necessary, and what the most general and simplest rules might be. In addition to offering greater clarity on the nature of the monotiling's construction, we hope to facilitate a search for physical systems that might manifest Chair Tilings, or which might profit from being studied on them as other physical systems have on other aperiodic tilings~\cite{FlickerVanWezel2015,Flicker2018,ZaporskiFlicker2019,Boyle2020,Flicker2020,Lloyd2022,Singh2024PRX}. 

\begin{figure}[t]
    \centering
    \includegraphics[width=0.9\columnwidth]{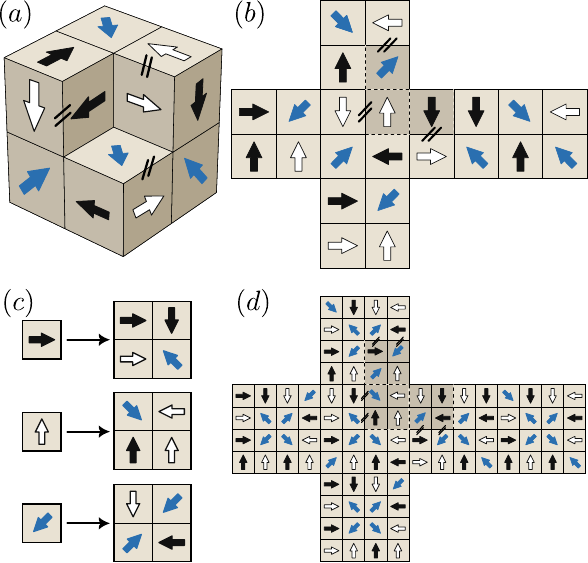}
\caption{(a) The Chair decorated with arrows. (b) A net of (a), in order to show the full decoration. Indented faces in (a) are dark faces in the net, and are reached across the edges marked with double black lines (stitches). Dashed lines are not edges on The Chair. (c) The decorations of Superchair44 are obtained from those of Chair44 by these composition rules. (d) The net of Superchair44.}
    \label{fig:arrows}
\end{figure}

To begin, we simplify the presentation of the matching rule of Ref.~\onlinecite{tsiokos} (Fig.~\ref{fig:DC}) using a set of three arrows (Fig.~\ref{fig:arrows}a). A square may meet another square only if their arrows align, and if the meeting introduces black-to-white or blue-to-blue. The black and white arrows align with square edges, while the blue arrows are diagonal. While the rules in Ref.~\onlinecite{tsiokos} have more than three inequivalent square decorations, our simplified arrow rules turn out to be equivalent, in the sense that they force all the same local (and therefore global) configurations~\footnote{In an earlier version we claimed our arrow rule was \emph{inequivalent} to the original, on the basis that squares can meet in different ways. We thank C.~Goodman-Strauss and J.~Walton for pointing out that the usual convention is to say that matching rules are equivalent if they force identical configurations around vertices, edges, and faces. We now adopt this nomenclature.}. Similar orientational rules were recently used to build a hexagonal monotile with a similar 2-adic hierarchy~\cite{MampustiWhittaker2020,WaltonWhittaker}. Our arrow construction immediately allows a generalisation to any geometric face decoration encoding exactly these constraints. 

In Fig.~\ref{fig:arrows}b we show a net of our arrow-decorated Chair44: squares sitting on cube faces form the obvious net (lighter tiles); the three indented faces are reached across the edges marked with double black lines (stitches), which also appear in Fig.~\ref{fig:arrows}a. The dashed lines on the net are not edges on The Chair.

\begin{figure}[t]
    \centering
    \includegraphics[width=0.9\columnwidth]{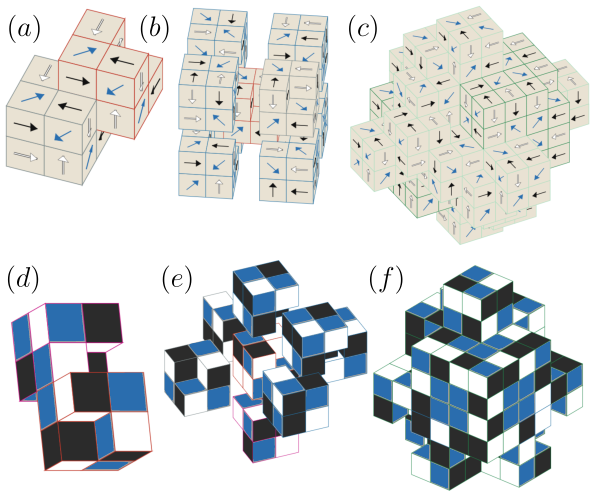}
\caption{Example outputs of our exact enumeration code. The top row shows arrow decorations, the bottom row face-colour decorations. Red lines: original Chair; blue lines: members of red's Superchair; magenta lines: mirror image; green lines: forced second layer. (a) One of the 44 ways for two Chair44s to meet. (b) One of the ways to enclose a Chair44 in a single shell (expanded for clarity). This example is a Superchair44 plus one Chair44 to complete it. (c) A 2-layer-deep shell. (d) One of the 611 ways for face-colour-decorated Chairs to meet; this example contains a mirrored Chair44. (e) An example of a 1-layer-deep shell, expanded; this one is illegal, meaning it will not admit a second shell, owing to the mirror image Chair. (f) A 2-layer-deep shell. Our accompanying code enumerates all cases and renders them as 3D models in html~\href{https://felixflicker.github.io/The_Chair/gallery.html}{available here}.}
    \label{fig:options}
\end{figure}

Using our arrow rules we verified that the steps of the proof outlined in Ref.~\onlinecite{tsiokos} continue to hold. In the supplementary material we provide a code~\footnote{Code \href{https://github.com/felixflicker/The_Chair}{available here}.} which enumerates all possible ways for two Chairs to meet. We find that there are 2388. Adding the arrow matching rules reduces this to 44, agreeing with the result of Ref.~\onlinecite{tsiokos} (we confirmed that they are the same 44). It should be noted that only 30 of these can appear in the infinite tiling. Our code generates an html file which allows the outputs to be viewed and rotated in 3D~\footnote{Gallery \href{https://felixflicker.github.io/The_Chair/gallery.html}{available here}.}; an example screenshot is shown in Fig.~\ref{fig:options}a.

We then enumerated all possible 1-layer-deep clusters of arrow-decorated Chairs around a central Chair44, and confirmed that there are 33. An example is shown in Fig.~\ref{fig:options}b. We then enumerated all possible ways to add a second layer around the first. We confirmed that only 15 of the original 33 are legitimate, and that all of these contain an identical Superchair around the central Chair44, and that all Chairs belong to only one Superchair. An example second-layer tiling is shown in Fig.~\ref{fig:options}c.

Inspecting the unique Superchair44, we find that the arrows on its surface are uniquely determined by those on Chair44. We show the composition rules in Fig.~\ref{fig:arrows}c, and the resulting net of Superchair44 in Fig.~\ref{fig:arrows}d. This completes the proof of strong aperiodicity, since Superchair44 must now obey exactly the same rules as Chair44, and hence by induction an infinite hierarchy results; this is known to be incompatible with periodicity or weak aperiodicity~\cite{GoodmanStrauss1998}.

The matching rules provide an additional constraint on Chair44 relative to undecorated Chairs. A 2D Chair Tiling can be built from (up to) four infinite supertiles meeting along a line or at a point~\cite{Vereshchagin2023}. Without decorations the two halves of a tiling divided by such a line can be slid relative to one another. Some matching rules continue to permit slides: translating the upper half of a Robinson tiling along a fault line preserves the rules but produces a tiling that cannot be generated by substitution~\cite{Robinson1971,GJS2012,Labbe2021}\footnote{We thank J.~Walton for pointing this out.}. The arrow matching rules of Chair44 force squares to meet squares; but it could still be the case that a discrete slide plane could be constructed. We eliminate this possibility as follows. The Superchair44 matching rules force slides to be multiples of 2 squares; the Super-Superchair matching rules require all slides to be multiples of 4, and so on. Hence the only admissible slide is the zero vector. 

Next we asked whether the arrow rules might be relaxed. To this end, we redid the exhaustive enumerations using only the colours of the arrows: a square is coloured either black, white, or blue, according to its arrow's colour, and the only restrictions are that black squares can only meet white, and vice versa, and blue can only meet blue. This removes any orientational constraints.

Of the 2388 ways for Chairs to meet, we now find that there are 611 ways for face-colour-decorated Chairs to meet, rather than the 44 possibilities with arrows. Following the nomenclature we therefore term this Chair611. One example is shown in Fig.~\ref{fig:options}c. 310 of these involve a mirrored Chair. The number of one-layer-deep shells rises from 33 to 49: the 16 new ones each contain a mirrored Chair. Hence the face-colouring rules of Chair611 are inequivalent to the matching rules of Chair44. However, when we enumerate the possible shells-of-shells, we find that no tilings with mirror images are legitimate. Remarkably, we find that only the same 15 shells of Chair44 ever appear in infinite tilings. Hence, three face colours alone suffice to force the aperiodic tilings.

\emph{Discussion}--Our primary motivation for generalising and simplifying the matching rules identified in Ref.~\onlinecite{tsiokos} was to facilitate physical realisations of Chair Tilings. Whereas the original rule involved small square pyramids of precise shapes and locations on the faces, in fact any geometrical structure or matching rule with the same symmetries will do. We found that simpler rules based on face colours also suffice. These ought to be simpler to implement physically. Our face colouring rules are notably less constraining, in that they allow many more neighbouring pairs and 1st level shells. It is fascinating that by the second level shells they have already shown that they must continue onto exactly the same infinite tilings. 

\begin{figure}[t]
    \centering
    \includegraphics[width=0.9\columnwidth]{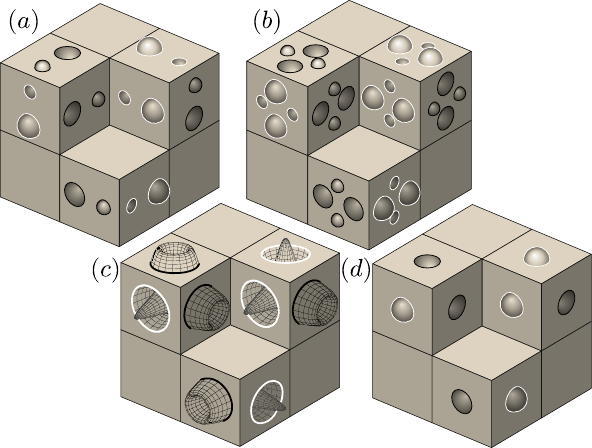}
\caption{Simple examples of geometrical matching rules that force aperiodicity. (a) An equivalent to the arrow rule of Fig.\ref{fig:arrows}a. (b),(c),(d) rules captured by the face colouring decorations of Chair611. (c,d) suggested by J.~Walton; (d) allows blue to meet black, but any such tiling fails to fill space owing to the resulting voids.}
    \label{fig:examples}
\end{figure}

We show geometric examples of our rules in Fig.~\ref{fig:examples}. Fig.~\ref{fig:examples}a shows a geometrical encoding of our arrow rules for Chair44, equivalent to Fig.~\ref{fig:DC}a. Figs.~\ref{fig:examples}b--d show encodings of the face colouring rules of Chair611. Fig.~\ref{fig:examples}b allows two orientations per face. Figs.~\ref{fig:examples}c,d have full rotational symmetry about a face; they were proposed by J.~Walton in response to an early draft of this pre-print. While the decorations in Fig.~\ref{fig:examples}d allow blue to meet black, this leaves a void and therefore does not constitute part of any legal tiling. For finite-size physical implementations, such as 3D printing, something equivalent to Fig.~\ref{fig:examples}c might be easier to work with.

Given the minimality of the constraints, it might not be too unreasonable to hope that Chair Tilings are realised somewhere in nature. Lock-and-key colloids, in which a spherical particle binds to a matching dimple~\cite{Sacanna2010}, and cubic colloids~\cite{Rossi2011}, could be promising candidates. Non-geometrical constraints, such as electrostatic interactions, could also work, suggesting possibilities for artificial constructions such as DNA self-assembly~\cite{Ke2012}, which has already produced the 2-adic Sierpinski triangle~\cite{Rothemund2004}. Arbitrary three-dimensional arrays of atoms can now be assembled in optical tweezers~\cite{Barredo2018}, and three-dimensional micromagnetic arrays can be made by two-photon lithography~\cite{Saccone23}. The vertices of Chair Tilings are a subset of the vertices of a simple cubic lattice. Hence it may also be possible to realise the tiling using interacting cold atoms in optical lattices, which have previously been used to create interacting quantum systems based on quasicrystals~\cite{Schneider1,Schneider2}.

The face colourings of Chair611 can also be used to define a lattice model by placing a 24-state variable on each site of a cubic lattice so as to record the orientation of the Chair, with nearest-neighbour interactions penalising illegal contacts; the ground states should exactly match Chair Tilings. Lattice models with limit-periodic ground states have been studied for the Taylor--Socolar tile: on slow cooling the ground state emerges through an infinite sequence of phase transitions, while a rapid quench produces a glass~\cite{ByingtonSocolar2012}; a three-dimensional model on the face-centred-cubic lattice with only nearest-neighbour interactions and only limit-periodic ground states is known, with first-order transitions~\cite{Marcoux2014} along with its diffraction pattern and domain-wall dynamics. The Chair is a natural cubic counterpart. It is interesting to ask whether it freezes hierarchically, and what its glassy states look like.

It is natural to ask whether there might be even simpler matching rules than our face colourings. We conjecture that there are not. For example, allowing blue squares to meet squares of any colour produces more than twenty thousand one-layer clusters instead of 49; merging black with white while retaining the arrow directions produces 6400. We did not enumerate second-level clusters, so it is possible that these or other simpler rules might work.

As to the question of what physical properties might be expected from Chair Tilings, fortunately much is already understood. For instance, Lee and Moody already identified that, placing a scattering centre (such as an atom) at the same position in each Chair of a given orientation, the resulting diffraction pattern must consist purely of sharp Bragg peaks~\cite{LeeMoody2001}. The distribution of these peaks is a clear experimental signature. The diffraction of limit-periodic structures, including the period-doubling sequence and the 2D Chair, has been computed explicitly~\cite{BaakeGrimm2011,BaakeGrimm2012}; the 2-adic internal space also appears in the recent translation-aperiodic tile of Greenfeld and Tao~\cite{GreenfeldTao2024}.

Face colourings of the squares have also been studied previously~\cite{FlomBenAbraham2022}. As with 2D Chair Tilings, for which matching rules that force aperiodicity were already known~\cite{Mozes1989,GoodmanStrauss1999}, the tiling is limit periodic and can be constructed using cut-and-project with a 2-adic internal space~\cite{BMS98,LeeMoody2001,BaakeGrimm2011}. This means, for example, that there is no underlying irrational number in Chair Tilings, in contrast to quasicrystals or the 2D aperiodic monotiles~\cite{hat,spectre} which are Pisot-type substitutions~\cite{Socolar2023,BaakeGahlerSadun2025,BaakeGahlerMazacMitchell2025,BaakeGahlerMazacSadun2025}.

Chair Tilings bear some similarities to the 2D aperiodic monotiles: a Superchair has at its centre a Chair that touches all other members of the Superchair, and which is essentially isolated from the rest of the tiling by those members (although there is a small opening, the red Chair in Fig.~\ref{fig:DC}b). This is similar to the Antihats in Hat Tilings~\cite{hat}, or Antispectres in Spectre Tilings~\cite{spectre,Singh24}. In physical models placed onto those tilings, these isolated central tiles play a key role, for example localising electronic wavefunctions at $\pi$ flux via quantum interference in tight binding models~\cite{Schirmann24}, caging fractionalised excitations in magnetic arrays~\cite{Tianyue26}, or forming a basis for quantum and classical degrees of freedom in exact solutions to quantum many-body problems~\cite{Singh24}. We hope that by identifying simple rules for enforcing Chair Tilings we might facilitate similar studies in 3D.

\emph{Note added}: a pre-print by Chaim Goodman-Strauss~\cite{GoodmanStrauss} appeared on the arXiv on the same day as ours; it identifies another arrow-based rule equivalent to that in Ref.~\onlinecite{tsiokos}.

\emph{Acknowledgments}--F.F. acknowledges support from the Engineering and Physical Sciences Research Council, Grant No.~EP/X012239/1. We thank I.~Tsiokos for sharing the results of Ref.~\onlinecite{tsiokos} prior to arXiving, and helpful discussions with C.~Goodman-Strauss, C.~S.~Kaplan, J.~Walton, D.~Rust, J.~Elliott, and R.~Moat which occurred after our initial arXiv listing. AI usage: we acknowledge Anthropic's Claude for coding the exhaustive numerical enumerations, and for useful discussions.

\bibliographystyle{apsrev4-2}
\bibliography{chair}

\end{document}